\documentclass{article}
\usepackage{spconf}

\usepackage{tikz,mathpazo}
\usetikzlibrary{shapes.geometric, arrows}
\usepackage{amsmath}
\usepackage{amssymb}
\usepackage{graphicx}
\usepackage{caption}
\usepackage{enumerate}
\usepackage{cite}
\usepackage[colorlinks=True]{hyperref}
\usepackage[latin9]{inputenc}
\usepackage{amssymb}
\usepackage{subcaption}
\usepackage[english]{babel}
\usepackage{algorithm}
\usepackage[noend]{algpseudocode}
\usetikzlibrary{shapes,arrows}
\usepackage{tikz}
\usetikzlibrary{arrows.meta}
\usetikzlibrary{arrows}
\usetikzlibrary{arrows,shapes,chains}

\title{
End-to-end Models with auditory attention in Multi-channel Keyword Spotting
}

\name{\Large Haitong Zhang \qquad Junbo Zhang \qquad Yujun Wang }

\address{\Large Xiaomi Inc., Beijing, China \\
\{zhanghaitong, zhangjunbo, wangyujun\}@xiaomi.com}

\begin{document}
\topmargin=0mm

\maketitle

\begin{abstract}

In this paper, we propose an attention-based end-to-end model for multi-channel keyword spotting (KWS), which is trained to optimize the KWS result directly. As a result, our model outperforms the baseline model with signal pre-processing techniques in both the clean and noisy testing data. We also found that multi-task learning results in a better performance when the training and testing data are similar. Transfer learning and multi-target spectral mapping can dramatically enhance the robustness to the noisy environment. At 0.1 false alarm (FA) per hour, the model with transfer learning and multi-target mapping gain an absolute 30\% improvement in the wake-up rate in the noisy data with SNR about -20.

\end{abstract}

\begin{keywords}
attention-based, end-to-end, multi-channel keyword spotting, single/multi-target spectral mapping, transfer learning
\end{keywords}

\section{INTRODUCTION}

Keyword spotting is a task to detect a pre-defined keyword from a continuous stream of speech. KWS recently has drawn increasing attention since it is used as a wake-up word on mobile devices.  In this case, the KWS model should satisfy the requirement of high accuracy, low-latency, and small-footprint.

The methods based on large vocabulary continuous speech recognition system (LVCSR) are used to process the audio offline \cite{miller2007rapid}. They generate rich lattices and search for the keyword.  Due to high-latency, these methods are not suitable for the mobile devices.  Another competitive technique for KWS is the keyword/filler Hidden Markov Model (HMM) \cite{rose1990hidden}. HMMs are trained separately for the keyword and non-keyword segments. At runtime, a Viterbi searching is needed to search for the keyword, which can be computationally expensive due to the HMM typology.

With the success of the application of neural network (NN) in automatic speech recognition (ASR), both keyword and non-keyword audio segments are used to train the same acoustic NN-based model. For example, in the Deep KWS model proposed by \cite{chen2014small}, a single DNN model is used to output the posterior probability of the sub-segment of the keyword and to make the KWS decision based on a confidence score using a posterior smoothing. To improve the performance, the more powerful neural networks such as convolutional neural network (CNN) \cite{sainath2015convolutional} and recurrent neural network (RNN) \cite{sun2016max} are used to substitute DNN. Inspired by these models, some end-to-end models have been proposed to directly output the probability of the whole keyword instead of sub-word, without any searching method or posterior handling \cite{bai2016end, arik2017convolutional, shan2018attention}.

Although tremendous improvement has been made, the previous models mainly focus on single-channel KWS. In industry, people usually use the microphone array for more complicated situations. Thus some signal processing techniques should be applied to convert the multi-channel signal into single-channel. However, these pre-processing techniques are sub-optimal because they are not optimized towards the final goal of interest\cite{seltzer2008bridging}. There are extensive literature in learning useful representations for multi-channel input in speech recognition. For example, \cite{liu2014using} concatenates the multi-channel signal into the network input. In  \cite{renals2014neural}, CNN is used to implicitly explore the spatial relationship between multiple channels. Attention-based methods have also been proposed to model the auditory attention in multi-channel speech recognition \cite{kim2015recurrent}.

Inspired by the application of attention mechanism in ASR \cite{kim2015recurrent}, we propose an attention-based end-to-end model for multi-channel KWS. Compared with \cite{kim2015recurrent}, our attention mechanism is computationally cheaper, which is more suitable in the KWS task. Transfer learning and multi-target spectral mapping are incorporated in the model to achieve an better result in the noisy evaluation data.

We describe the proposed model in Section 2. The experiment data, setup, and results follow in section 3. Section 4 closes with the conclusion.

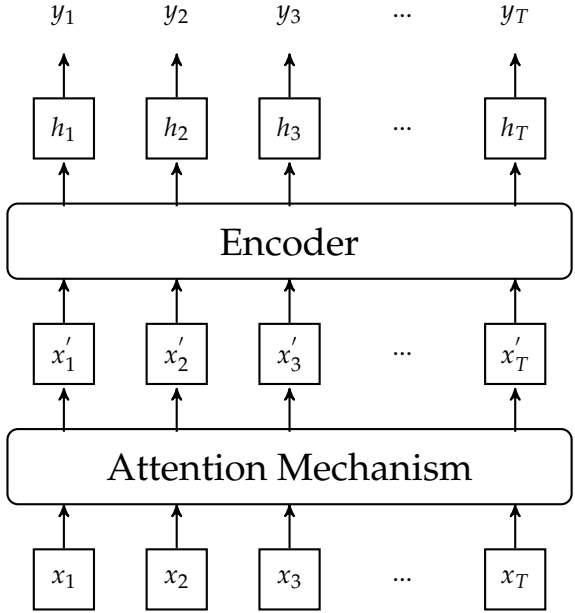
\begin{figure}
\tikzstyle{startstop} = [rectangle, rounded corners, minimum width = 3cm, minimum height = 0.7cm, text centered, draw = black]
\tikzstyle{startstop1} = [rectangle,  minimum width = .5cm, minimum height = .5cm, text centered, draw = black]
\tikzstyle{startstopw} = [rectangle,  minimum width = .7cm, minimum height = .9cm, text centered, draw = white]
\tikzstyle{startstop2} = [rectangle, rounded corners, minimum width = 7.5cm, minimum height = 1cm, text centered, draw = black]
\tikzstyle{startstop2w} = [rectangle, rounded corners, minimum width = 7.5cm, minimum height = 1cm, text centered, draw = white]
\tikzstyle{startstop2wl} = [rectangle, rounded corners, minimum width = 5.5cm, minimum height = 1cm,align=left, draw = white]
\tikzstyle{startstop2p} = [rectangle, rounded corners, minimum width = .8cm, minimum height = .9cm, text centered, draw = white]
\tikzstyle{io} = [trapezium, trapezium left angle = 30, trapezium right angle = 150, minimum width = 3cm, text centered, draw = black, fill = white]
\tikzstyle{io2} = [trapezium, trapezium left angle = 30, trapezium right angle = 150, minimum width = 2.5cm, draw = black, fill = white]
\tikzstyle{io3} = [trapezium, trapezium left angle = 30, trapezium right angle = 150, minimum width = 2cm, draw = black, fill = white]
\tikzstyle{process} = [rectangle, minimum width = 3cm, minimum height = 1cm, text centered, draw = black]
\tikzstyle{decision} = [diamond, minimum width = 3cm, minimum height = 1cm, text centered, draw = black]
\tikzstyle{arrow} = [thick, -, >= stealth]
\tikzstyle{arrow2} = [thick, ->, >= stealth]

\begin{centering}
\begin{tikzpicture}[->,>=stealth',shorten >=1pt,auto,node distance=1.5cm,
                    thick,main node/.style={rectangle,draw,minimum width=.8cm, minimum height=.8cm}]

  \node[main node] (1) {$x_1$};
  \node[main node] (2) [right of=1] {$x_2 $};
  \node[main node] (3) [right of=2] {$x_3$};
  \node[startstopw] (4) [right of=3] {$...$};
  \node[main node] (5) [right of=4] {$x_T$};

  \node[startstop2p] (1p) [above of = 1]{};
  \node[startstop2p] (2p) [above of = 2]{};
  \node[startstop2p] (3p) [above of = 3]{};
  \node[startstop2p] (4p) [above of = 4]{};
  \node[startstop2p] (5p) [above of = 5]{};

  \node[startstop2] (6) [above of=3] {\Large Attention Mechanism};

  \node[main node] (9) [above of=6] {$x^{'}_3$};
  \node[main node] (8) [left of= 9] {$x^{'}_2 $};
  \node[main node] (7) [left of =8]{$x^{'}_1$};
  \node[startstopw] (10) [right of=9] {$...$};
  \node[main node] (11) [right of=10] {$x^{'}_T$};

  \node[startstop2p] (7p) [above of = 7]{};
  \node[startstop2p] (8p) [above of = 8]{};
  \node[startstop2p] (9p) [above of = 9]{};
  \node[startstop2p] (10p) [above of = 10]{};
  \node[startstop2p] (11p) [above of = 11]{};

  \node[startstop2] (12) [above of = 9] {\Large Encoder};

    \node[main node] (13) [above of=12] {$h_3$};
  \node[main node] (14) [left of= 13] {$ h_2 $};
  \node[main node] (15) [left of =14]{$h_1$};
  \node[startstopw] (16) [right of=13] {$...$};
  \node[main node] (17) [right of=16] {$h_T$};

      \node[startstopw] (18) [above of=13] {$y_3$};
  \node[startstopw] (19) [left of= 18] {$ y_2 $};
  \node[startstopw] (20) [left of =19]{$y_1$};
  \node[startstopw] (21) [right of=18] {$...$};
  \node[startstopw] (22) [right of=21] {$y_T$};

  \draw[->] (1) -- (1p);
  \draw[->] (2) -- (2p);
  \draw[->] (3) -- (3p);
  \draw[->] (5) -- (5p);
  \draw[->] (1p) -- (7);
  \draw[->] (2p) -- (8);
  \draw[->] (3p) -- (9);
  \draw[->] (5p) -- (11);
  \draw[->] (7) -- (7p);
  \draw[->] (8) -- (8p);
  \draw[->] (9) -- (9p);
  \draw[->] (11) -- (11p);
    \draw[->] (7p) -- (15);
  \draw[->] (8p) -- (14);
  \draw[->] (9p) -- (13);
  \draw[->] (11p) -- (17);
     \draw[->] (15) -- (20);
  \draw[->] (14) -- (19);
  \draw[->] (13) -- (18);
  \draw[->] (17) -- (22);

\end{tikzpicture}
\end{centering}
\caption{The proposed model architecture.}
\label{model}
\end{figure}

\section{The Proposed Model}
As illustrated in Fig.\ref{model}, the model mainly consists of three components: (i) the attention mechanism, (ii) the sequence-to-sequence training, (iii) the decoding smoothing.
\subsection{Attention Mechanism}
The attention mechanism we use is the soft attention, as proposed in\cite{chowdhury2017attention}.  For each time-step, we compute a 6-dimensional attention weight vector $A_t^{ch}$ as followed:

\begin{gather}
A^{ch}_t = softmax(V_t * tanh(Wx_t^{ch}+b))   \\
x_t' = \sum_{c=1}^{ch} {A_t^j x_t^j}
\end{gather}

where $x_t^{ch}$ is a 6$\times$40 input feature matrix, W is a 40$ \times$128 weight matrix, b is a 128-dimension bias vector, and $V_t$ is a 128-dimension vector. A softmax function is applied for normalization. $x_t'$ is the weighted sum of the  multi-channel inputs $x_t^{ch}$.

\subsection{Sequence to sequence training}




The training framework we use is sequence-to-sequence. The encoder learns the higher representations $h$ for the enhanced speech features $x'$. With a linear transformation and a softmax function, the probability of the whole keyword can be predicted at each frame. \vspace{1pt}


\begin{figure}[h]
 \centering
\begin{subfigure}{0.5\textwidth}
\includegraphics[width=0.8\textwidth, height=0.1\textheight]{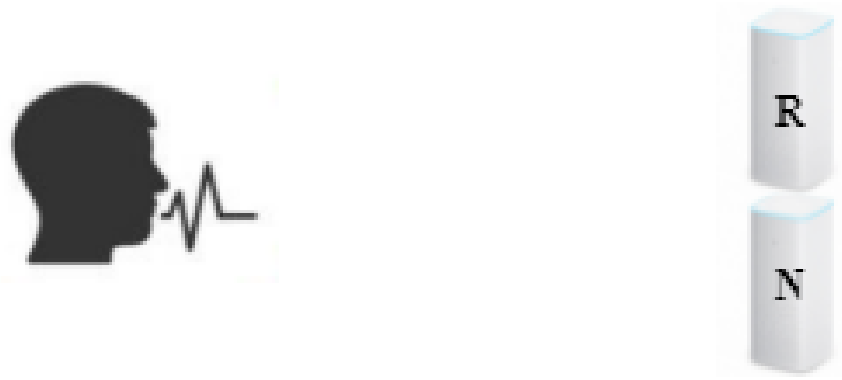}
\caption{The first situation for recording.}
\label{fig:record1}
\end{subfigure}
\begin{subfigure}{0.5\textwidth}
\includegraphics[width=0.8\textwidth, height=0.06\textheight]{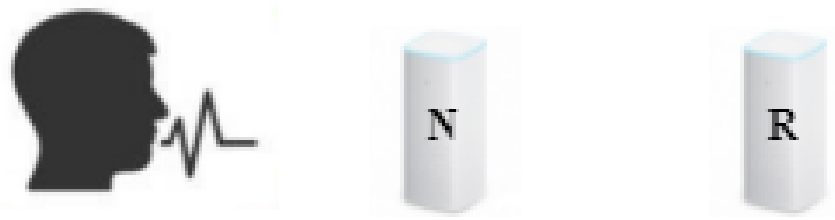}
\caption{The second situation for recording.}
\label{fig:record2}
\end{subfigure}

\caption{Two situations for recording the noisy testing data in a 4m*4m*3.5m labatory, where N denotes noise and R denotes recording device.}
\label{fig:record12}
\end{figure}

\noindent\textbf{\bf Multi-task Learning.} \label{multitask}
To improve the performance, we take the spectral mapping as an auxiliary task for our KWS model. The model learns the nonlinear mapping between the multi-channel speech features and the single-channel speech features, which is inspired by \cite{chen2015speech}. The target speech features come from the traditional signal processing techniques. Although we depreciate the idea of separating the front-end signal processing techniques and the acoustic model, we conjecture whether the multi-task framework can improve the performance. The loss function is:
\begin{equation}
\centering Loss_{total} = \alpha * Loss_{KWS} + (1-\alpha) Loss_{Map\_clean}
\end{equation}

\noindent\textbf{\bf Transfer Learning.} We also adopt transfer learning to improve the performance in the noisy environment. Transfer learning \cite{pan2010survey} refers to initializing the model parameters with the corresponding parameters of a trained model. Here we initialize the network using the proposed multi-channel KWS model trained with the relatively clean data, and fine-tune the model with only noisy data.\vspace{1pt}

\noindent\textbf{\bf Multi-target Mapping.} Since it is difficult to train the model with all noisy data, we propose multi-target spectral mapping. We conjecture that with more mapping targets, the spectral mapping can converge better than learning the nonlinear relationship between the noisiest input and the cleanest output. Compared with the spectral mapping mentioned above, two extra mapping targets are involved when training (detailed in sub-section \ref{data}). The loss function is described as followed:

\begin{equation}
\begin{split}
    Loss_{total} & = \alpha * Loss_{KWS} + \beta * Loss_{Map\_clean}\\
    &+ \theta * Loss_{Map\_noise1}  + \delta * Loss_{Map\_noise2}
\end{split}
\end{equation}

\noindent with the constraint that $\alpha + \beta + \theta + \delta = 1$.

\subsection{Decoding}

When decoding, our model, takes as the input a $6*40$ feature matrix and outputs the keyword spotting probability at each frame. We adopt a posterior probability smoothing method, and finally the decision is made based on the average probability of $n$ frames.


\section{Experiments}

\subsection{Datasets}\label{data}

The training data consists of 240k utterances of the keyword (which includes 120k ones with echo and the other without echo), and 200 hours of negative examples, with 10\% of them used for validation. The evaluation data includes 50 hours of filler data and 48k keyword data (with 50\% echo keywords and 50\% non-echo ones). We also record 1k noisy keywords as Fig.\ref{fig:record12} illustrates. They consists of two equal parts, which are recorded in two situations. The first half (referred to hard-noisy data, with the average SNR about -20) is recorded as shown in Fig.\ref{fig:record1} where the recording device is close to the music noise and the speaker is 3 meters away from the device. The other (referred to easy-noisy data, with the average SNR about -18) are recorded as Fig.\ref{fig:record2} shows. The distance between the device and speaker remains unchanged, but the between the device and the music source is 1 meter.

Besides that, we also recorded 50 hours of music for the multi-target mapping experiment. In the experiment, we randomly add music to the 120k non-echo keywords and 200 hours of negative examples to generate the noisy training data. Algorithm \ref{algorithm} shows the procedure of creating multiple mapping targets.

\subsection{Experiment setups}\label{setup}
In the baseline single-channel KWS model, the front-end component mainly includes beamforming and acoustic echo cancellation (AEC). These two blocks are constructed as proposed in \cite{grondin2013manyears} and \cite{valin2007adjusting}, respectively.

The input feature in all the experiment is the trainable PCEN \cite{wang2017trainable}. The 40-dimension filter-bank features are extracted using a window of 25ms with a shift of 10ms. The encoder in the experiment is two GRU (Gated Recurrent Units) layers \cite{chung2014empirical}  and one fully-connected layer. Both the GRU and FC layer have 128 units, with a 0.9 dropout rate. In the multi-task models, two tasks use two separate FC layers. Adam optimizer \cite{kingma2014adam} is used to update the training parameters, with the batch size and initial learning rate is 64 and 0.001, respectively. The $\alpha, \beta, \theta, and \delta $ value in the spectral mapping is 0.5, 0.2,0.2 ,and 0.1, respectively, since they are reasonably good in the development data.

\begin{figure*}[ht]

\begin{subfigure}{0.5\textwidth}
\includegraphics[width=1\linewidth]{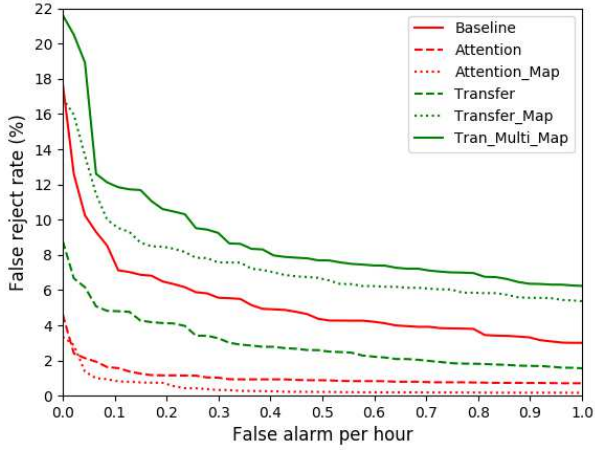}
\caption{ROC for the Non-echo data}
\label{fig:no_echo}
\end{subfigure}
\begin{subfigure}{0.5\textwidth}
\includegraphics[width=1\linewidth]{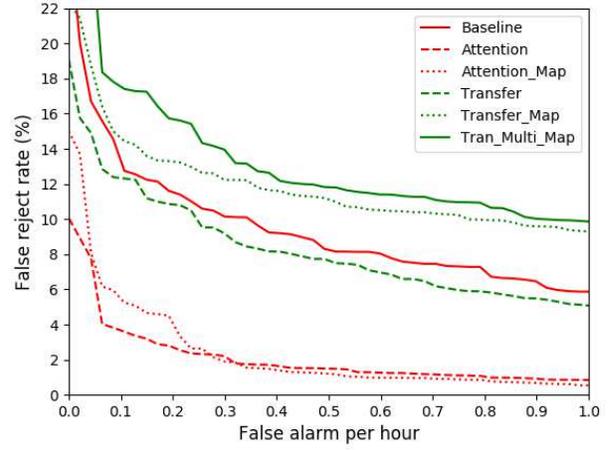}
\caption{ROC for the Echo data}
\label{fig:echo}
\end{subfigure}\\
\begin{subfigure}{0.5\textwidth}
\includegraphics[width=1\linewidth]{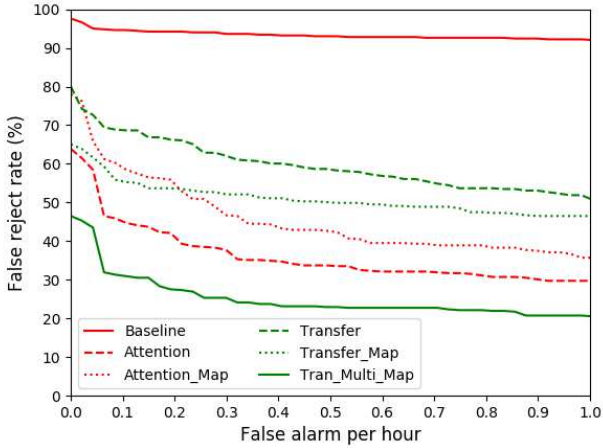}
\caption{ROC for the Easy-noisy data}
\label{fig:easy}
\end{subfigure}
\begin{subfigure}{0.5\textwidth}
\includegraphics[width=1\linewidth]{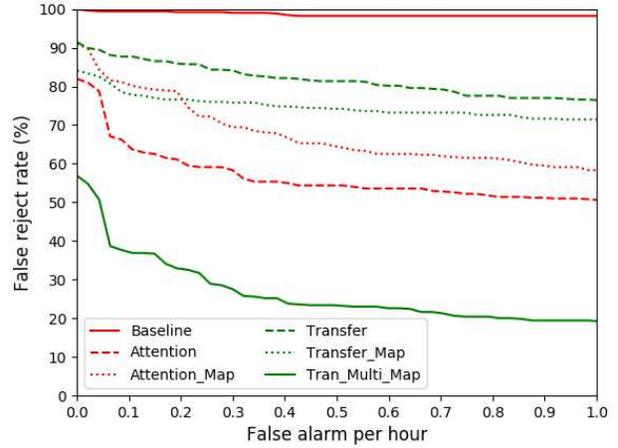}
\caption{ROC for the Hard-noisy data}
\label{fig:hard}
\end{subfigure}

\caption{The results of the baseline model and the proposed models, with the smoothing frame $n=12$.}
\label{result_baseline}
\end{figure*}

\begin{algorithm}
\caption{Procedure for creating noisy training data in the multi-target spectral mapping}\label{algorithm}
\begin{algorithmic}[1]
\For{each multi-channel wav a in all wavs}
\State select one music clip b randomly
\State add b into a, with SNR is about -10 \Comment{Input}
\State convert a into single-channel c \Comment{Target 1 }
\State convert b into single-channel d
\State add d into c, with SNR is about +5\Comment{Target 2}
\State add d into c, with SNR is about +10 \Comment{Target 3}
\State \textbf{return} Input, Target 1, Target 2, Target 3
\EndFor
\end{algorithmic}
\end{algorithm}

\subsection{Impact of Attention mechanism}

We first evaluate the performance of the baseline model and the proposed multi-channel models. The baseline model uses the signal processing techniques in sub-section \ref{setup}. The input of the signal processing techniques is seven channels, while the input of the proposed model (i.e. Attention) is only six channels, without the reference signal for AEC.

It is obvious that Attention outperforms the baseline model in all the evaluation data sets. At 0.5 false alarm (FA) per hour, Attention gains an absolute 4\% improvement and 7\%, respectively in the non-echo and  echo data (Fig. \ref{fig:no_echo} and Fig.\ref{fig:echo}).

The difference in the performances becomes larger in the noisy data (Fig.\ref{fig:easy} and Fig.\ref{fig:hard}). The performance improvements are 40\% and 60\%, respectively in the hard-noisy data and easy-noisy data. This great difference may be largely attributed to that the signal processing techniques are not robust to the noisy environment, especially when the noise is close to the wake-up device.

\subsection{Impact of multi-task learning}
As indicated in Fig.\ref{fig:no_echo} and Fig.\ref{fig:echo}, the proposed model with spectral mapping (i.e. Mapping) outperforms Attention slightly, which to some degree confirms our conjecture. However, the result gets worse in the noisy data (Fig.\ref{fig:easy} and Fig.\ref{fig:hard}). Such a difference lie in the difference between the training data and noisy testing data.

\subsection{Impact of transfer learning and multi-target mapping}
To increase the noise-robustness of the model, we initialize the model with the parameters of Attention and fine-tune the model with the artificial noisy training data (detailed in sub-section \ref{data}). As shown in Fig.\ref{fig:no_echo} and Fig.\ref{fig:echo}, all the models with transfer learning (i.e. Transfer, Transfer\_map, and Tran\_Multi\_Map) perform worse than Attention in both non-echo and echo testing data. The reason lies in the difference between the training data and the testing data. However, the main target of the transfer learning and multi-target mapping is the noisy data. As illustrated in Fig.\ref{fig:easy} and Fig.\ref{fig:hard}, transfer learning and single-target spectra mapping do not result in a better result than Attention, which confirms the difficulties in training the model with only noisy data. However, the model with the transfer learning and multi-target mapping (i.e. Tran\_Multi\_Map) outperforms all the models by a large margin in the noisy data. At 0.5 false alarm per hour, Tran\_Multi\_map gains an absolute 30\% and 10\% improvement over Attention, respectively in the hard\_noisy data and easy\_noisy data.

\section{CONCLUSIONS}

Without the reference signal for AEC, the proposed attention-based model for multi-channel KWS out-performs the baseline model in all the testing data. With spectral mapping, the performance can gain a slight improvement when the training data and testing data are similar. In addition, transfer learning and multi-target spectral mapping can enhance the model's robustness to the noisy environment, which shed lights on the NN-based speech enhancement in ASR.


\section{Acknowledge}
The authors would like to thanks the colleagues from the acoustic group for useful discussion.



\begin{thebibliography}{10}

\bibitem{miller2007rapid}
David~RH Miller, Michael Kleber, Chia-Lin Kao, Owen Kimball, Thomas Colthurst,
  Stephen~A Lowe, Richard~M Schwartz, and Herbert Gish.
\newblock Rapid and accurate spoken term detection.
\newblock In {\em Eighth Annual Conference of the International Speech
  Communication Association}, 2007.

\bibitem{rose1990hidden}
Richard~C Rose and Douglas~B Paul.
\newblock A hidden markov model based keyword recognition system.
\newblock In {\em Acoustics, Speech, and Signal Processing, 1990. ICASSP-90.,
  1990 International Conference on}, pages 129--132. IEEE, 1990.

\bibitem{chen2014small}
Guoguo Chen, Carolina Parada, and Georg Heigold.
\newblock Small-footprint keyword spotting using deep neural networks.
\newblock In {\em Acoustics, Speech and Signal Processing (ICASSP), 2014 IEEE
  International Conference on}, pages 4087--4091. IEEE, 2014.

\bibitem{sainath2015convolutional}
Tara~N Sainath and Carolina Parada.
\newblock Convolutional neural networks for small-footprint keyword spotting.
\newblock In {\em Sixteenth Annual Conference of the International Speech
  Communication Association}, 2015.

\bibitem{sun2016max}
Ming Sun, Anirudh Raju, George Tucker, Sankaran Panchapagesan, Gengshen Fu,
  Arindam Mandal, Spyros Matsoukas, Nikko Strom, and Shiv Vitaladevuni.
\newblock Max-pooling loss training of long short-term memory networks for
  small-footprint keyword spotting.
\newblock In {\em Spoken Language Technology Workshop (SLT), 2016 IEEE}, pages
  474--480. IEEE, 2016.

\bibitem{bai2016end}
Ye~Bai, Jiangyan Yi, Hao Ni, Zhengqi Wen, Bin Liu, Ya~Li, and Jianhua Tao.
\newblock End-to-end keywords spotting based on connectionist temporal
  classification for mandarin.
\newblock In {\em Chinese Spoken Language Processing (ISCSLP), 2016 10th
  International Symposium on}, pages 1--5. IEEE, 2016.

\bibitem{arik2017convolutional}
Sercan~O Arik, Markus Kliegl, Rewon Child, Joel Hestness, Andrew Gibiansky,
  Chris Fougner, Ryan Prenger, and Adam Coates.
\newblock Convolutional recurrent neural networks for small-footprint keyword
  spotting.
\newblock {\em arXiv preprint arXiv:1703.05390}, 2017.

\bibitem{shan2018attention}
Changhao Shan, Junbo Zhang, Yujun Wang, and Lei Xie.
\newblock Attention-based end-to-end models for small-footprint keyword
  spotting.
\newblock {\em arXiv preprint arXiv:1803.10916}, 2018.

\bibitem{seltzer2008bridging}
Michael~L Seltzer.
\newblock Bridging the gap: Towards a unified framework for hands-free speech
  recognition using microphone arrays.
\newblock In {\em Hands-Free Speech Communication and Microphone Arrays, 2008.
  HSCMA 2008}, pages 104--107. IEEE, 2008.

\bibitem{liu2014using}
Yulan Liu, Pengyuan Zhang, and Thomas Hain.
\newblock Using neural network front-ends on far field multiple microphones
  based speech recognition.
\newblock In {\em Acoustics, Speech and Signal Processing (ICASSP), 2014 IEEE
  International Conference on}, pages 5542--5546. IEEE, 2014.

\bibitem{renals2014neural}
Steve Renals and Pawel Swietojanski.
\newblock Neural networks for distant speech recognition.
\newblock In {\em Hands-free Speech Communication and Microphone Arrays
  (HSCMA), 2014 4th Joint Workshop on}, pages 172--176. IEEE, 2014.

\bibitem{kim2015recurrent}
Suyoun Kim and Ian Lane.
\newblock Recurrent models for auditory attention in multi-microphone distance
  speech recognition.
\newblock {\em arXiv preprint arXiv:1511.06407}, 2015.

\bibitem{chowdhury2017attention}
FA~Chowdhury, Quan Wang, Ignacio~Lopez Moreno, and Li~Wan.
\newblock Attention-based models for text-dependent speaker verification.
\newblock {\em arXiv preprint arXiv:1710.10470}, 2017.

\bibitem{chen2015speech}
Zhuo Chen, Shinji Watanabe, Hakan Erdogan, and John~R Hershey.
\newblock Speech enhancement and recognition using multi-task learning of long
  short-term memory recurrent neural networks.
\newblock In {\em Sixteenth Annual Conference of the International Speech
  Communication Association}, 2015.

\bibitem{pan2010survey}
Sinno~Jialin Pan, Qiang Yang, et~al.
\newblock A survey on transfer learning.
\newblock {\em IEEE Transactions on knowledge and data engineering},
  22(10):1345--1359, 2010.

\bibitem{grondin2013manyears}
Fran{\c{c}}ois Grondin, Dominic L{\'e}tourneau, Fran{\c{c}}ois Ferland, Vincent
  Rousseau, and Fran{\c{c}}ois Michaud.
\newblock The manyears open framework.
\newblock {\em Autonomous Robots}, 34(3):217--232, 2013.

\bibitem{valin2007adjusting}
Jean-Marc Valin.
\newblock On adjusting the learning rate in frequency domain echo cancellation
  with double-talk.
\newblock {\em IEEE Transactions on Audio, Speech, and Language Processing},
  15(3):1030--1034, 2007.

\bibitem{wang2017trainable}
Yuxuan Wang, Pascal Getreuer, Thad Hughes, Richard~F Lyon, and Rif~A Saurous.
\newblock Trainable frontend for robust and far-field keyword spotting.
\newblock In {\em 2017 IEEE International Conference on Acoustics, Speech and
  Signal Processing (ICASSP)}, pages 5670--5674. IEEE, 2017.

\bibitem{chung2014empirical}
Junyoung Chung, Caglar Gulcehre, KyungHyun Cho, and Yoshua Bengio.
\newblock Empirical evaluation of gated recurrent neural networks on sequence
  modeling.
\newblock {\em arXiv preprint arXiv:1412.3555}, 2014.

\bibitem{kingma2014adam}
Diederik~P Kingma and Jimmy Ba.
\newblock Adam: A method for stochastic optimization.
\newblock {\em arXiv preprint arXiv:1412.6980}, 2014.






\end{thebibliography}

\end{document}